\documentclass[hyper]{JHEP3}

\pdfoutput=1
\usepackage{epsfig}
\usepackage{latexsym}
\usepackage{amsfonts}
\usepackage{amsmath}
\usepackage{amsthm}
\usepackage{amssymb}
\usepackage{amsbsy}
\usepackage{multirow}

\newcommand\rmi{\ensuremath{\rm i}}
\newcommand\ra[1]{\ensuremath{\xrightarrow{#1}}}

\newcommand\nn{\nonumber \\{}}

\newcommand{\cp}[1]{\ensuremath{\mathbb{CP}^{#1}}}
\newcommand{\wpr}[2]{\ensuremath{W\mathbb{CP}^{#1}_{#2}}}
\newcommand\curlo{\ensuremath{\mathcal{O}}}
\newcommand\curli{\ensuremath{\mathcal{I}}}
\newcommand\curle{\ensuremath{\mathcal{E}}}
\newcommand\curlf{\ensuremath{\mathcal{F}}}
\newcommand\curleb{\ensuremath{\mathcal{E}^\bullet}}
\newcommand\curlfb{\ensuremath{\mathcal{F}^\bullet}}
\newcommand\curlgb{\ensuremath{\mathcal{G}^\bullet}}
\newcommand{\dercat}[1]{\ensuremath{\mathsf{D}\left({#1}\right)}}
\def\be{\begin{equation}}
\def\ee{\end{equation}}
\def\bea{\begin{eqnarray}}
\def\eea{\end{eqnarray}}
\def\bi{\begin{itemize}}
\def\ei{\end{itemize}}
\def\ba{\begin{align}}
\def\ea{\end{align}}

\def\Df{\ensuremath{\rm D4}}
\def\Ds{\ensuremath{\rm D6}}
\def\Dzb{\ensuremath{\rm \overline{D0}}}
\def\Dtb{\ensuremath{\rm \overline{D2}}}
\def\Dsb{\ensuremath{\rm \overline{D6}}}
\DeclareMathOperator{\rank}{rank}
\DeclareMathOperator{\Hom}{Hom}
\DeclareMathOperator{\imaginary}{Im}
\DeclareMathOperator{\td}{td}

\title{\begin{center} Exact indices for D--particles, flow trees and the derived category \end{center}}
\author{\begin{center}{Walter Van Herck}\end{center}\\
\begin{center}{Institute for Theoretical Physics}\\
{K.~U.~Leuven }\\
{Celestijnenlaan 200D}\\
{B-3001 Leuven, Belgium}\end{center}
\bigskip
\centerline{{\rm E-mail}: \email{waltervh@itf.fys.kuleuven.be}}}

\abstract{In \cite{VanHerck:2009ww} D--particle degeneracies were calculated using novel methods. By analysing this refined computation of the index in the setting of the topological B model, a correspondence is found between flow trees in the supergravity description and B--branes in the topological theory. This enables us to use the same refinement also for single flows. A concrete example will be given, providing strong support for this method, while at the same time confirming the prediction of a new elliptic genus in \cite{VanHerck:2009ww}.}


\preprint{KUL-TF-10/05}

\begin{document}
\section{Introduction}\label{sect:intro}
The BPS sector of type IIA string theory compactified on a Calabi Yau threefold has many interesting features. In the low energy limit, the theory reduces to a $d=4$, $\mathcal{N}=2$ supergravity theory, where, for large enough charges, the BPS states represent supersymmetric black hole solutions. As in the case of Seiberg--Witten theory, $\mathcal{N}=2$ supersymmetry seems to be just enough supersymmetry to enable one to derive much of the properties of the theory, while at the same time exhibiting physically interesting features. The main property explored in this paper is the degeneracy of BPS states of fixed charge.

In the supergravity theory, the BPS states correspond in general to, possibly multicentered, black holes. For small charges however, we will refer to the states as D--particles, in analogy with the pointlike D--branes that are wrapped on internal cycles of the Calabi Yau manifold in the type II string theory. The attractor mechanism is a very powerful tool in analyzing these solutions. For multicentered solutions, the use of this mechanism culminated in the establishment of split flow trees, conjectured to be an existence criterion of BPS states in the full string theory \cite{Denef:2007vg}.

These same states can also be studied as supersymmetric states in a two dimensional sigma model, with the Calabi Yau threefold as a target space. This is a result of the fact that the index is protected by supersymmetry, allowing one to consider the limit of zero string coupling. By performing a topological twist of this theory \cite{Witten:1988xj, Witten:1991zz}, one obtains a topological field theory in two dimensions, which still contains the BPS states under consideration. In the topological B model, they correspond to B--branes and can be represented as objects in the derived category of coherent sheaves. This description provides a very powerful mathematical framework for the analysis of supersymmetric branes and their bound states on a Calabi Yau manifold.

In \cite{Collinucci:2008ht, VanHerck:2009ww}, the degeneracy of D--particles for some specific one modulus Calabi Yau threefolds was calculated using split flow trees. A refinement of the index calculation was proposed in \cite{Collinucci:2008ht} and further worked out in \cite{VanHerck:2009ww}. The present paper will be concerned of extending this framework also to single flow states. In the course of doing this, further support for the refined index calculation is provided\footnote{By \emph{refined index computation}, we mean a refined method for calculating the indices of specific charge states and not a calculation of a refined index.}.

The organization of this paper is as follows. Section \ref{sect:dercat} will provide the necessary background and notation for B--branes and the derived category. It is followed by a short recapitulation of a refined index calculation of \cite{VanHerck:2009ww}, while section \ref{sect:refinedwc} will consider this example from the perspective of the derived category. In the same section, a conjectured correspondence between flow trees and objects in the derived category will be put forth. These statements are then verified in a concrete example in section \ref{sect:casestudy}. Finally, a short discussion and outlook is provided in section \ref{sect:discussion}.

\section{B--branes and the derived category}\label{sect:dercat}
The material in section \ref{sect:refinedwc} relies heavily on the description of topological B--branes as objects in the derived category of coherent sheaves. The present section will therefore set our notation and give some examples of B--branes. More details of this framework can be found in the excellent review article of Aspinwall \cite{Aspinwall:2004jr}.

\subsection*{General B--branes}
As already conjectured in \cite{Kontsevich:1994}, B--branes can be described as objects in the derived category of coherent sheaves $\dercat X$. The objects in this category are complexes of coherent sheaves. The need for coherent sheaves, instead of locally free sheaves, arises from the fact that we will need the cohomologies of these complexes, something which is really only well defined in an abelian category. Thus, the category of locally free sheaves is enlarged by adding (co)kernels inside the category of $\curlo_X$ modules. An example of such a B--brane is:
\be
\dotsb \ra{} 0 \ra{} \curle^{-1} \ra{d_{-1}} \curle^0 \ra{d_0} \curle^1 \ra{d_1} 0 \ra{} \dotsb \, ,
\ee
where $\curle^i$ denotes a coherent sheaf on $X$ and the $d_i$ are morphisms obeying $d_{n+1}\circ d_n=0$. Following the notation in the literature, this complex will be denoted as \curleb .

The morphisms in the derived category consist of chain maps between the complexes, which are a set of $f_n: \curle^n \ra{} \curlf^n$, such that ever square commutes (actually, the morphisms are chain maps modulo chain homotopy, but this will not interfere with our discussion here). On top of these morphisms, one defines the notion of a quasi--isomorphism: chain maps that induce an isomorphism between the cohomologies of the complexes. In such a case, one also adds their inverse as a morphism. Complexes linked by quasi--isomorphisms are then considered to be isomorphic. The set of (coherent) sheaf complexes and these morphisms then form the derived category of B--branes.

In the next subsections, examples of objects in the derived category we will encounter are discussed. To shorten the notation we will indicate a brane of dimension $p$ as $\rm Dp$ and the corresponding anti--brane as $\rm \overline{Dp}$, revealing our eventual interest in describing D--branes.

\subsection*{6--branes with flux}
The representation of (anti--)6--branes with $U(1)$--flux as a sheaf complex is rather straightforward. The sheaf is locally free and corresponds to a holomorphic line bundle. In homogeneous coordinates, the transition functions can be represented by a homogeneous degree $d$ polynomial. In this case, the corresponding sheaf will be denoted by $\curlo(d)$. If the manifold, on which the sheaf is defined, needs to be specified, we will denote it in a subscript, e.g. $\curlo_X(d)$.

Note that the complex
\be
0 \ra{} 0 \ra{} \curlo \ra{Id_\curlo} \curlo \ra{} 0 \, ,
\ee
with $Id_\curlo$ the identity morphism, is exact, indicating a quasi--isomorphism between the zero complex and $\curlo \ra{Id_\curlo} \curlo$. The two 6--branes can thus annihilate each other. This observation suggests that anti--branes are represented by the same complex, but shifted by one position. This turns out to be correct, so the anti--brane of $\curleb$ is $\curleb[1]$, where $\curleb[n]$ indicates the complex $\curleb$ shifted by $n$ places to the left.

\subsection*{Adding 0--branes}
Adding an \Dzb\ to a \Ds\ can be done in the following way. In a neighborhood of the position of the \Dzb , we use the inhomogeneous complex coordinates $(x,y,z)$ and define a morphism from $\curlo^{\oplus 3}$ to $\curlo$ by $(f_1,f_2,f_3)\ra{x,y,z}xf_1+yf_2+zf_3$. The cokernel object of this morphism is called the skyscraper sheaf and has support equal to the origin. For a more general point $p\in X$, denote this sheaf as $\curlo_p$. The \Ds\ plus \Dzb\ at the point $p$ can then be represented as the following complex:
\be
\dotsb \ra{} 0 \ra{} \curlo \ra{} \curlo_p \ra{} 0 \ra{} \dotsb \, ,
\ee
where the only non--trivial morphism takes a holomorphic function to its value in $p$. In terms of the coordinate ring, the polynomial ring $\mathbb{C}[x,y,z]$ gets quotiented out by the maximal ideal generated by the three linear functions $(f_1,f_2,f_3)$. This leaves the ring $\mathbb{C}$, representing constant functions, exactly as one would have expected for the function ring over a point.

By adding the kernel of this morphism and its object, one can construct a short exact sequence:
\be
0 \ra{} \curli_p \ra{} \curlo \ra{} \curlo_p \ra{} 0 \, ,
\ee
where $\curli_p$ is called the ideal sheaf in $p$. This short exact sequence implies a quasi--iso--morphism between $\curli_p$ and $\curlo \ra{} \curlo_p$. So we may equally well consider the ideal sheaf as representing a \Ds\ with one \Dzb .

The construction of 6--branes with more \Dzb 's is very similar. To include two \Dzb 's, instead of considering three linear functions mapping $\curlo^{\oplus 3}$ to $\curlo$, we now take one quadratic function and two linear functions. For simplicity, we take the morphism defined by multiplication with $x^2-a^2$, $y$ and $z$ respectively. The support of the cokernel sheaf will then be restricted to the two points $(a,0,0)$ and $(-a,0,0)$. For $a\neq 0$, the sheaf looks like two isolated points with fiber $\mathbb{C}$. When $a=0$ however, we are faced with a single point as support and a coordinate ring different from $\mathbb{C}$ (quotienting out by the three functions, we are left with functions of the form $ag+b$, with $a,b\in\mathbb{C}$ and $g$ a linear function, determined, up to a constant factor, by the three defining functions). Physically, it is very natural to include these states as multiple branes on the same position. Mathematically, the inclusion of multiplicities will force us to use the notion of a scheme, rather than a variety. In algebraic geometry, this is also closely related to the notion of a blow--up procedure: putting two points on the same locus, the degrees of freedom are the locus of these points plus the direction from which they approach each other (in three dimensions, the direction, or as we have seen, the extra linear polynomial lives in $\cp{2}$). The reader who is unfamiliar with the language of schemes, can consult \cite{HartshorneAG} or just think of these as describing (sub)varieties with multiplicity.

\subsection*{Adding 2--branes}
Finally, we will look at \Ds\Dtb\Dzb\ states. As an example, take a curve defined by the zero locus of the ideal generated by the functions $x$ and $y$ (as in the previous examples, we work in local coordinates $(x,y,z)$). This is just the $z$--axis. To this curve we add a point, defined by the three functions $x$, $y-a$ and $z$. The union of these two varieties will be defined as the zero locus of the intersection of the two aforementioned ideals. This new ideal will be generated by the functions $x$, $y(y-a)$ and $yz$. The coordinate ring consists of the direct sum of polynomial functions in $z$ (denoted $\mathbb{C}[z]$) and the constant functions $\mathbb{C}$. This is directly related to the regular functions on the curve and the point. If $a\ra{} 0$, the point will be located on the curve. As a variety, the zero locus is just the curve. But as a scheme, the point is not `lost', as we can see from the coordinate ring. This ring consists of $\mathbb{C}[z]\oplus\mathbb{C}\cdot\epsilon$, with $\epsilon^2=0$. In general, the coordinate ring consists of the direct sum of the coordinate ring on the curve and $\mathbb{C}$ times a linear function that is normal to the curve. The blow--up procedure in this case thus includes a $\cp{1}$ of normal directions. 

\section{Refined index calculation}\label{sect:recap}
In this section, the results from \cite{VanHerck:2009ww} will be briefly stated, giving extra attention to one special case. This will allow us to understand the details of this calculation from the perspective of the derived category of B--branes in section \ref{sect:refinedwc}. For more details about the refined calculation reviewed below, we refer the reader to \cite{VanHerck:2009ww}.

\subsection{General calculation strategy}
To calculate the degeneracy of BPS states in Calabi--Yau compactifications of type IIA string theory, in \cite{VanHerck:2009ww} we used the modularity of the elliptic genus (a formal power series, generating the indices for states of variable electric and fixed magnetic charge). Knowledge of a finite number of terms, called polar terms, is then sufficient to determine the whole power series \cite{Dijkgraaf:2000fq, deBoer:2006vg, Manschot:2007ha}. The same polar terms in principle describe BPS states that are realized as multicentered configurations in the low energy effective supergravity theory \cite{Denef:2007vg}. The polar state degeneracies are then calculated as follows:
\bi
\item For a fixed (polar) charge, find all possible split flow trees.
\item For each flow tree, calculate the degeneracy as the product of the indices of the different centers with the index of the tachyonic states between them. As the tachyon fields may perceive special configurations of the lower dimensional D--brane charges in the centers differently, a refinement may be necessary to achieve the correct answer.
\item Sum up all these indices for a given charge.
\ei
Without refinement, the index for a fixed charge vector would then become:
\be
\Omega (\Gamma) = \sum_{\Gamma_1 +\Gamma_2 =\Gamma}(-1)^{\langle\Gamma_1,\Gamma_2 \rangle -1} \langle\Gamma_1 ,\Gamma_2 \rangle \Omega (\Gamma_1)\Omega (\Gamma_2) \, ,
\ee
where $\Omega(\Gamma)$ is the index for the BPS state with charge vector $\Gamma$, the summation is over different split flow trees for a fixed total charge $\Gamma$ and $(-1)^{\langle\Gamma_1,\Gamma_2 \rangle -1} \langle\Gamma_1 ,\Gamma_2 \rangle$ is the index of the tachyon field between the constituent centers.

\subsection{Example on the octic}
As our test case, we revisit the degree $8$ hypersurface $X$ in the weighted projective space \wpr{4}{11114}, with homogeneous coordinates $(x_1,x_2,x_3,x_4,x_5)$. As the defining equation for the Calabi--Yau hypersurface, we take:
\be
p_{octic} = x_5^2 + p^{(8)}(x_1,x_2,x_3,x_4) = 0 \, .
\ee
Note that the first term is needed to avoid the presence of the point $(0,0,0,0,1)$, which is a $\mathbb{Z}_2$ orbifold singularity. The homogeneous degree $8$ polynomial $p^{(8)}$ is chosen such that $p_{octic}=0$ and $dp_{octic}=0$ has no common solution. In this case, $p_{octic}$ is called a transverse polynomial. Define $H\in H^2(\wpr{4}{11114})$ as the Poincar\'e dual of the homology class of the hypersurface $x_1=0$. With slight abuse of notation, $H$ will also denote its pullback to the Calabi--Yau $X$ and its Poincar\'e dual in $H_4(X)$. The total Chern class of $X$ is then
\be
c(X)=1+22H^2-148H^3
\ee
and $\int_X H^3 = \int_{\wpr{4}{11114}}8H^4 = 2$. Integration of the top Chern class thus gives the Euler characteristic $\chi(X)=-296$.

Our focus will now turn to the calculation of the refined index for the state with \Df --charge $H$ and two added \Dzb --branes. For this state we found only one split flow tree, consisting of a \Ds\ with flux $H$ and two added \Dzb 's and a \Dsb . As stated in \cite{VanHerck:2009ww}, the tachyon field is identified by a section of the class $H$ line bundle, so we can write
\be
T = a_1x_1+a_2x_2+a_3x_3+a_4x_4 \, .
\ee
The moduli space of such sections is just $\cp{3}$, as a global scaling of the coefficients $a_i$ by a factor $\lambda\in\mathbb{C}^*$ can be absorbed in the scaling of the homogeneous coordinates $x_i$. As an extra constraint, the tachyon field had to vanish on the locus of the added \Dzb 's. So in general, this reduces the moduli space from $\cp{3}$ to $\cp{1}$. The number of constraints is more specifically 
\be
\rank \begin{pmatrix} x_1 & x_2 & x_3 & x_4 \\
y_1 & y_2 & y_3 & y_4 \end{pmatrix} \, ,
\ee
where $x_i$ and $y_i$ denote the homogeneous coordinates of the two \Dzb 's. Constraint loss thus occurs when $y_i = \lambda x_i$ for $i = 1, \dotsc , 4$. For $x_i$ not all zero, which would not constitute a point of $X$ anyway, these equations can be recast as three independent homogeneous degree one equations. In general, they would have $\int_X H^3 =2$ solutions, which are geometrically identified with $y_i = \lambda x_i$ for $i = 1, \dotsc , 4$ and $y_5 = \pm \lambda^4 x_5$. When $x_5=0$, the only non--general situation, only one solution remains: the \Dzb 's sit at the same location.

When the \Dzb 's are at the same location, a blow--up needs to be performed and the tachyon map additionally needs to vanish in this direction. The number of constraints then becomes
\be
\rank \begin{pmatrix} x_1 & x_2 & x_3 & x_4 \\
X_1 & X_2 & X_3 & X_4 \end{pmatrix} \, ,
\ee
with $X\equiv \sum_{i=1}^5 X_i\partial_i$ in the tangent space of $X$ at the \Dzb\ locus and corresponding to the blow--up direction. As calculated in \cite{VanHerck:2009ww}, when $x_5=0$, a constraint loss is encountered in the blow--up direction $X_5\partial_5$.

If we denote by $N_{total}$ the index of all configurations of the \Ds\ with two \Dzb 's, and by $N_{cl}$ the index of the configurations where a constraint loss occurs, we can calculate the index of the state of interest as:
\be
\Omega = (N_{total} - N_{cl})\cdot \chi(\cp{1}) + N_{cl} \cdot \chi(\cp{2}) \, ,
\ee
since the index of the \Dsb\ is just one. Also note that $\chi(\cp{n})$ is the Euler characteristic of $\cp{n}$, denoting the relevant index for the tachyon field.

In \cite{VanHerck:2009ww} we identified $N_{total}$ with the Donaldson--Thomas invariant $N_{DT}(0 ,2)$ , where $N_{DT}(\beta ,n)$ `counts' subschemes $Z$ of $X$, with $[Z]=\beta\in H_2(X)$ and $\chi (\curlo_Z) = n$. Furthermore, we put $\mathcal{N}^{(g)}_{DT}(0,2)\equiv N_{total} - N_{cl}$ and $\mathcal{N}^{(s)}_{DT}(0,2)\equiv N_{cl}$ such that $\mathcal{N}^{(g)}_{DT}(0,2) + \mathcal{N}^{(s)}_{DT}(0,2) = N_{DT}(0,2)$.

\section{Refined indices in the derived category}\label{sect:refinedwc}
The example from the previous section will now be analyzed from the derived category point of view. This will already reveal some generic features of the correspondence between coherent sheaves and flow trees. In the last part of this section, this correspondence will be stated and commented on. There, we will also conjecture that it can be used to calculate exact degeneracies of states with fixed charges, even in the presence of single flows.

\subsection{\Ds --\Dsb\ bound states in the derived category}
As explained in section \ref{sect:dercat}, a single \Ds\ with flux $H$ and two \Dzb 's can be represented by the sheaf complex
\be \label{eq:d6htwod0complex}
\dotsb \ra{} 0 \ra{} \curlo(1) \ra{\psi} \curlo_{p_1p_2} \ra{} \dotsb \, ,
\ee
where $\psi$ is defined as the map that takes a section of $\curlo(1)$ to its values on $p_1$ and $p_2$.

The \Dsb\ can be viewed as the following complex
\be \label{eq:d6bcomplex}
\dotsb \ra{} \curlo \ra{} 0 \ra{} 0 \ra{} \dotsb \, .
\ee
In this case, the bound state of these two objects will be
\be \label{eq:d6d6btwod0complex}
\dotsb \ra{} \curlo \ra{\phi} \curlo(1) \ra{\psi} \curlo_{p_1p_2} \ra{} \dotsb \, ,
\ee
with $\phi \in \Hom(\curlo ,\curlo(1))$ a morphism between $\curlo$ and $\curlo(1)$. The morphisms $\phi$ are in one--to--one correspondence with degree $1$ homogeneous functions $f$ on $X$, with $\phi$ equivalent to multiplication by $f$. The functions $f$ are thus of the form
\be
f = a_1x_1 + a_2x_2 + a_3x_3 + a_4x_4 \, .
\ee
Since (\ref{eq:d6d6btwod0complex}) must be a complex, we have the constraint $\psi\circ\phi = 0$. This directly translates into the requirement that $f$ vanishes on $p_1, p_2$. When the \Dzb 's are in the same location $p$, the map $\psi$ takes a section of $\curlo(1)$ to its value in $p$ and its derivative at $p$ in the direction of the blow--up. The constraints we put on the tachyon field in section \ref{sect:recap} thus have a clear interpretation in the derived category.

In retrospect, we see that an index of a specific flow tree can be calculated in the derived category as the index of possible configurations of a sheaf complex. In the current example, the only components of our complex with a non--trivial index are the locations of the \Dzb 's (which immediately fix the morphism $\psi$) and the possible morphisms $\phi$. Since the set of possible morphisms $\phi$ depends on the exact locations of the \Dzb 's, this clarifies the refined index computation of \cite{VanHerck:2009ww}. Also note that the indices of the complexes (\ref{eq:d6htwod0complex}) and (\ref{eq:d6bcomplex}) are just $N_{DT}(0,2)$ and $1$ respectively.

\subsection{Stability}
The description of B--branes as objects in the derived category of coherent sheaves does not itself provide for the notion of stability. One has to supplement the category with extra stability data to incorporate this. Stability issues follow from the fact that BPS conditions for D--branes in the untwisted two--dimensional sigma model are stronger than the conditions one imposes on B--branes in the twisted topological field theory.

The criteria for $\Pi$--stability of B--branes \cite{Douglas:2000gi, Douglas:2000ah, Douglas:2000qw, Fiol:2000wx, Aspinwall:2001dz} will be briefly stated so as to understand how stability will be treated further on. First, one defines a grading on sheaf complexes
\be
\xi (\curleb ) \equiv \frac{1}{\pi}\arg Z(\curleb ) \pmod 2 \, ,
\ee
with $Z(\curleb )$ the holomorphic central charge of the B--brane $\curleb$. Note that $Z$ depends on the complexified K\"ahler structure $B+\rmi J$ of the Calabi--Yau manifold. On top of this, we demand that this grading varies continuously over the complexified K\"ahler moduli space as long as the state is stable (no grading is defined for unstable objects). One also has
\be
\xi (\curleb [n])= \xi (\curleb ) + n \, .
\ee
$\Pi$--stability is then stated as follows. For a distinguished triangle (which is a generalization of short exact sequences in the case of non--abelian categories)
\be
\curleb \ra{e} \curlfb \ra{f} \curlgb \ra{g} \curleb [1] \, ,
\ee
with $\curleb$ and $\curlfb$ stable, $\curlgb$ is stable with respect to decay into $\curleb [1]$ and $\curlfb$ if and only if $\xi (\curlfb ) < \xi (\curleb ) + 1$.

An important remark regarding this notion of stability is that these conditions still do not fix the set of stable B--branes. One needs to specify the stable B--branes and their gradings at a basepoint in moduli space\footnote{To be precise, one needs to fix it at a basepoint in the Teichm\"uller space, which is a finite covering space of the moduli space.}. In this paper, we only consider degeneracies of BPS states at large K\"ahler structure, so this means we can fix the set of stable D--brane states by using arguments of the supergravity description of these states at large K\"ahler structure.

In \cite{Denef:2007vg} a necessary condition for the stability of a bound state of two charges $\Gamma_1$ and $\Gamma_2$ was stated as
\be \label{eq:intconstraint}
\langle \Gamma_1 , \Gamma_2 \rangle \imaginary (Z_1\bar{Z}_2)_{\infty} > 0 \, .
\ee
If we take $B+\rmi J \ra{} \rmi\infty$, the central charge is
\be
Z(\curleb ) \sim \int_X \rm{e}^{-(B+\rmi J)} ch(\curleb )\sqrt{\td (X)} \, ,
\ee
and for $\Gamma_1 = (p_1^0,p_1,q_1,q_{1,0})$ and $\Gamma_2 = (p_2^0,p_2,q_2,q_{2,0})$, we have that
\be \label{eq:z1z2bar}
\imaginary (Z_1\bar{Z}_2)_{\infty} \sim J^5 (p_1p_2^0-p_2p_1^0) + \curlo (J^3) \, .
\ee
By using equations (\ref{eq:intconstraint}) and (\ref{eq:z1z2bar}), the stability of a given bound state at $B+\rmi J \ra{} \rmi\infty$ can easily be verified.

\subsection{Flow trees and stable B--branes}
From the description of stability in the previous subsection, we will now argue that single flows can also be counted using the refined index counting technique. Suppose that a certain bound state reaches its attractor value before hitting the wall of marginal stability of its constituent states. In this case, the flow tree picture in supergravity would give a single flow. From the perspective of the topological field theory, it is however reasonable to say that the index of such a state will not change by varying the moduli in the whole stable region, which has the wall of marginal stability as its boundary. The index could then just as well be counted on this wall, where a \emph{partial} (due to the refined aspect of these indices) factorization occurs.

To make things less cumbersome, assume for now that the constituent states, or the decay products if you wish, have only one realization in the derived category. This means that for a fixed charge of a decay product, there will only be one (stable) isomorphism class of sheaf complexes having this charge. The correspondence between the split flow tree picture in supergravity and the derived category could then be stated as follows:
\bi
\item Each split flow tree (up to the equivalence defined by threshold walls) will correspond to exactly one isomorphism class of stable sheaf complexes in the derived category. This allows one to calculate, for each of these inequivalent trees, the index corresponding to this sheaf complex.
\item Each single flow tree can, in principle, consist of multiple equivalence classes of sheaf complexes in the derived category. Once all these classes have been identified and a representative is chosen for each of these classes, this again allows the computation of the total index as the sum of the individual indices of each class.
\ei
In the case where the constituent states can have multiple (non--isomorphic) representatives in the derived category, each single flow of the constituent charge, which is part of the total split flow tree, should be treated as in our second argument, which is as a genuine single flow tree.

To demonstrate this correspondence, and to give concrete support for it, in the next section we will revisit a specific index, which was incompletely calculated in \cite{VanHerck:2009ww}. Besides an error in that calculation, which will be rectified, the incompleteness was mainly due to the presence of a single flow tree. The present understanding of the correspondence between (split) flow trees and the derived category now allows to complete this calculation. Furthermore, the exact equality between the calculated index and the prediction from modular invariance gives very strong support to our arguments.

\section{Single flow indices: a case study}\label{sect:casestudy}
The calculation is this section is a completion, plus a minor correction, of the one performed in section 5 of \cite{VanHerck:2009ww}. The state of interest is a pure \Df\ with two added \Dzb 's on the degree ten hypersurface in \wpr{4}{11125}. This Calabi--Yau hypersurface $X$ has total Chern class: $c(X)=1+34H^2-288H^3$ and, since $\int_X H^3=1$, its Euler characteristic is $-288$. The defining equation in $\wpr{4}{11125}$ is
\be
p(x_i) = x_5^2+x_4^5+p^{(10)}(x_1,x_2,x_3) = 0 \, ,
\ee
with $x_i, \, i=1\ldots 5$ the homogeneous coordinates on $\wpr{4}{11125}$.

By calculating the indices of the polar states (pure \Df\ and \Df\ with one \Dzb ), we arrived at a new prediction of its elliptic genus\footnote{Without the refined calculation, one arrives at a different prediction for this elliptic genus.}
\be \label{eq:decanticgenus}
Z_0 = q^{-\frac{35}{24}}\left( 3 - 575q + 271'955q^2 + 206'406'410q^3 + 21'593'817'025q^4 + \dotsb \right)
\ee
To confirm this new prediction, we tried to verify the third term, which is the first non--polar term, in this genus. Since this charge state has a single flow however, we were unable to correctly verify its index. The calculation in this section will thus not only confirm the correctness of the correspondence between flow trees and the derived category, but will also provide extra support for the refined index calculation scheme by providing a non--trivial check on the correctness of a new prediction of an elliptic genus.

\subsection{A split flow}
Let us now turn our attention to the charge state of interest, a \Df\ with two added \Dzb 's. As already stated in \cite{VanHerck:2009ww}, we find a split flow tree with a fluxed \Ds\ and two \Dzb 's and a \Dsb . This state is very similar to the one we discussed in sections \ref{sect:recap} and \ref{sect:refinedwc}. The number of constraints on the tachyon field for two \Dzb 's at different location are
\be \label{eq:condecantic}
\rank \begin{pmatrix} x_1 & x_2 & x_3 \\
y_1 & y_2 & y_3 \end{pmatrix} \, ,
\ee
where $x_i$ and $y_i,\, i=1\ldots5$ denote the homogeneous coordinates of the two \Dzb 's.

The special point $(x_1,x_2,x_3)=(0,0,0)$ will be treated first and denoted by $X_{123}$. Its Euler characteristic is just $\chi_{123}=1$, as would be expected for a single point. If one of the \Dzb 's sits in $X_{123}$, the number of constraints is clearly just one, instead of two for general positions of the \Dzb 's. If both are in $X_{123}$, we have to perform a blow--up and it is easy to see that the full set of blow--up directions ($\cp{2}$) does give an extra constraint.

For the remaining set of points $X\setminus X_{123}$, suppose we fix a constraint by putting one \Dzb . From (\ref{eq:condecantic}), we see that constraint loss ($\rank <2$) occurs for a second \Dzb\ whose coordinates satisfy two degree one equations (whose coefficients are determined by $(x_1,x_2,x_3)$). Additivity of the Chern class then determines
\be \label{eq:cxparallel}
c(X_{cl}(x_i)) = \frac{c(X)}{(1+H)^2} = 1-2H \, ,
\ee
where $X_{cl}(x_i)$ denotes the locus with constraint loss (which depends on the coordinates $x_i$ of the first \Dzb ). Its index\footnote{Without further specification, the index of a variety or scheme will refer to its Euler characteristic.} is $\int_{X_{cl}} -2H = \int_X -2H^3 = -2$. However, the special point $X_{123}$, which we already treated in the previous paragraph, will always be a solution to (\ref{eq:cxparallel}), so the index of `parallel' solutions on $X\setminus X_{123}$ is $\chi_\parallel = -2 -1=-3$. This index counts the number of solutions for the first \Dzb\ in general position, so one has to check if special situations can occur. For $(y_1,y_2,y_3)$ fixed, up to scaling, the only possibility for this to happen would be when $p^{(10)}(y_1,y_2,y_3)=0$. In this case, one still has `$-3$ solutions' where $(y_4,y_5)\neq (0,0)$, but there is also an extra solution for which $(y_4,y_5)=(0,0)$. Since each point with $(y_4,y_5)=(0,0)$ will have an inequivalent set of degree one coordinates $(y_1,y_2,y_3)$, this locus counts the number of inequivalent classes of points that satisfy $p^{(10)}(y_1,y_2,y_3)=0$. Define $X_{45}$ as the locus $x_4=x_5=0$, then
\be
c(X_{45})= \frac{c(X)}{(1+2H)(1+5H)} = 1-7H \, ,
\ee
and $\chi_{45}\equiv \chi(X_{45})=\int_{X_{45}} -7H = \int_X 2H*5H*(-7H) = -70$. As each of these classes denotes a set with index $-2$, the total index of this set of points is $\chi_{special}\equiv (-70)(-2)=140$. For each of these, there are $\chi_\parallel +1 =-2$ solutions that result in a constraint loss.

Finally, one has to calculate the number of constraints in case of a blow--up in $X\setminus X_{123}$. Since $(x_1,x_2,x_3)\neq (0,0,0)$, one can fix one of these coordinates to $1$, meaning that constraint loss will only occur for tangent directions $X^i\partial_i$ with $X^1=X^2=X^3=0$. The condition for the direction to lie in the tangent space of the Calabi--Yau hypersurface then becomes
\be
2X^5x_5+5X^4x_4^4 = 0 \, .
\ee
If $(x_4,x_5)\neq (0,0)$, this gives one direction with constraint loss. In case $(x_4,x_5)=(0,0)$, there is a $\cp{1}$ of directions with constraint loss.

Now we have all we need to calculate the refined index of the \Ds $_H$--2\Dzb , \Dsb\ state. It has the following contributions without constraint loss:
\bi
\item $\frac{1}{2}\left[ (\chi -\chi_{123})^2 -(\chi -\chi_{123}-\chi_{special})\chi_\parallel - \chi_{special}(\chi_\parallel +1)\right]=41'257$: This counts the generic situation with two \Dzb 's in different location and giving two independent constraints. Note that $\chi -\chi_{123}$ is the index of $X\setminus X_{123}$.
\item $(\chi -\chi_{123} -\chi_{45})\left[ \chi(\cp{2})-1\right] + \chi_{45}\left[ \chi(\cp{2})-\chi(\cp{1})\right]=-508$: The index for a blow--up in $X\setminus X_{123}$, without constraint loss. The locus $X_{45}$ is dealt with separately, because of the enlarged set of directions with constraint loss (a $\cp{1}$).
\ei
These indices sum up to $40'749$.

The index contributions where constraint loss occurs are given by:
\bi
\item $\frac{1}{2}(\chi -\chi_{123}-\chi_{special})(\chi_\parallel -1) = 858$: This index denotes the situation where one \Dzb\ is in a generic location ($X\setminus (X_{123}\cup X_{special})$) and the other gives constraint loss. The `$-1$' in the last factor subtracts the situation where a blow--up needs to be performed.
\item $\frac{1}{2}\chi_{special}(\chi_\parallel +1 -1) = -210$: This index refers to a similar situation as in the previous item, but with an extra point giving constraint loss (hence, the `$+1$' in the last factor).
\item $\chi_{123}(\chi - \chi_{123}) = -289$: The index of the situation where one \Dzb\ has $(x_1,x_2,x_3)=(0,0,0)$ and the other sits in $X\setminus X_{123}$.
\item $(\chi -\chi_{123}-\chi_{45})\cdot 1 + \chi_{123}\cdot\chi(\cp{2}) + \chi_{45}\cdot\chi(\cp{1})= -356$: These are the blow--up situations with constraint loss. Three different cases are distinguished: `general' point, $x_1=x_2=x_3=0$ and $x_4=x_5=0$.
\ei
The index of the constraint loss situations totals $3$.

In the notation of \cite{VanHerck:2009ww}, we have
\be
N_{DT}(0,2)=\mathcal{N}^{(g)}_{DT}(0,2)+\mathcal{N}^{(s)}_{DT}(0,2)=40'749 + 3=40'752 \, ,
\ee
and the index of our specific state is
\be
\mathcal{N}^{(g)}_{DT}(0,2)\cdot\chi(\cp{0})+\mathcal{N}^{(s)}_{DT}(0,2)\cdot\chi(\cp{1}) = 40'752\cdot 1 + 3\cdot 2 = 40'755 \, .
\ee

\subsection{A single flow}
On top of the previous state, corresponding to a split flow tree, one also finds a \Ds\ with a degree one rational curve and flux $2H$ plus a \Dsb\ with flux $H$. The individual charges of these constituents are:
\begin{align}
\Gamma_1 & = (1,2,\frac{29}{12},\frac{7}{6}) \nn
\Gamma_2 & = (-1,-1,-\frac{23}{12},-\frac{19}{12}) \, ,
\end{align}
adding up to the total charge $\Gamma = (0,1,\frac{1}{2},-\frac{5}{12}) $.

This state corresponds to a single flow tree, as the flow in moduli space reaches its attractor value before crossing the wall of marginal stability. The identification of the two individual charges can thus be viewed solely in the picture of B--branes, where the sheaf complex will consist of a bound state of these two charges. Also note that the attractor point will never satisfy the stability condition of equation (\ref{eq:intconstraint}) for a split into $\Gamma_1$ and $\Gamma_2$, so stability in this case should be phrased in the more complex setting of $\Pi$--stability, as is appropriate in the derived category\footnote{Thanks to Frederik Denef for clarifying this to me.}.

To calculate its index, the extension of the refined index calculation is necessary. As stated before however, this does not affect the calculation itself very much, as this state could be moved in moduli space to the wall of marginal stability without changing its index. We are therefore in a position to calculate the index as in the case of split flows. 

The curve on the \Ds\ will completely fix the tachyon field (see \cite{VanHerck:2009ww} for a similar case on the octic), implying that the total index will equal the number of degree one rational curves, which can be found to be $231'200$ (see for example \cite{Huang:2006hq}).

Adding up the contributions from the split and the single flow, the result is $40'755+231'200=271'955$, which exactly matches the modular prediction in (\ref{eq:decanticgenus})! 

Furthermore, the Donaldson--Thomas invariant $\mathcal{N}_{DT}(1,1)=435'827$ consists of contributions from these degree one rational curves and degree one curves with genus one and two, with added \Dzb\ charges. This means that, by using only the contribution from the rational curves, this index is also a refined one.

Note that the modular prediction itself was also a result of an index refinement in \cite{VanHerck:2009ww}. The remarkable correspondence between our calculation and the modular prediction therefore provides strong support to both the ideas presented at the end of section \ref{sect:refinedwc} as to the new elliptic genus of \cite{VanHerck:2009ww}.

\section{Discussion}\label{sect:discussion}
In this paper, the refined index calculation method of \cite{Collinucci:2008ht, VanHerck:2009ww} was extended to the computation of single flow states. This was made possible through a conjectured correspondence between (split) flow trees in supergravity and B--branes in the derived category. The main idea behind this correspondence was the observation that indices of objects in the derived category should not jump when varying the moduli inside the region of stability, which is bounded by a wall of marginal stability. The fact that some flow trees end at their attractor values before hitting this wall, only indicates that in the supergravity description there is no decay of a `localized' D--particle into a bound state of constituent charges, sitting at different locations of spacetime.

The correspondence enables a refined calculation method for states of fixed charge, whether or not they correspond to split flows. For a specific charge state, one first finds all inequivalent stable sheaf complexes, each one corresponding to a single or split flow. Then the indices of these complexes are calculated, using a refined calculation method, which deals with the possible non--trivial fibration of the tachyon field moduli over the moduli space of ideal sheaves. Summing up all these contributions then gives the total index for the state of fixed charges. As demonstrated by a given example in section \ref{sect:casestudy}, the results are in exact agreement with the prediction from modular invariance of the elliptic genus.

It would be interesting to check if this method can also be applied for constituent charges supporting $U(N)$ gauge fields, thus corresponding to multiple \Ds\ or \Dsb\ branes. This would certainly render the computation more complicated, although, in principle, it should not invalidate the argumentation of the present paper. A more problematic issue is the apparent discrepancy between the indices found here and the degeneracies implied by the wall crossing formula of \cite{Denef:2007vg}. Although the origin of this discrepancy is still unclear, the exact agreement with the modular prediction at least strongly suggests the validity of our results in the regime where this modular invariant elliptic genus is valid. The analysis and possible solution of this issue will be left for future research.

\acknowledgments{I want to express my gratitude to Frederik Denef, Albrecht Klemm and Andres Collinucci for useful comments and enlightening discussions. This work is supported in part by the FWO - Vlaanderen, project G.0235.05 and in part by the Federal Office for Scientific, Technical and Cultural Affairs through the `Interuniversity Attraction Poles Programme -- Belgian Science Policy' P6/11-P.}

\newpage

\bibliographystyle{JHEP}
\bibliography{ref}

\end{document}